\documentclass[aps,prl,reprint, amsmath, amssymb,superscriptaddress]{revtex4-1}

\usepackage{bm}
\usepackage[retainorgcmds]{IEEEtrantools}
\usepackage{graphicx}
\usepackage{mathrsfs}
\usepackage{amsmath}
\usepackage{amssymb}
\usepackage{color}
\usepackage{amsfonts}
\usepackage{amsthm}
\usepackage{times,txfonts}
\usepackage{enumerate}
\usepackage{dsfont}
\usepackage{nicefrac}

\newcommand{\ud}{\,\mathrm{d}}

\DeclareMathOperator{\tr}{tr}
\DeclareMathOperator{\var}{Var}
\DeclareMathOperator{\cov}{Cov}

\newcommand{\trans}{^{\text{T}}}

\DeclareMathOperator{\csch}{csch}

\DeclareMathOperator{\arctanh}{arctanh}

\newlength{\eqboxstorage}

\begin{document}

\title{Thermodynamic uncertainty relations from exchange fluctuation theorems}
\date{\today}
\author{Andr\'e M. Timpanaro}
\email{a.timpanaro@ufabc.edu.br}
\affiliation{Universidade Federal do ABC,  09210-580 Santo Andr\'e, Brazil}
\author{Giacomo Guarnieri}
\email{guarnieg@tcd.ie}
\affiliation{Department of Physics, Trinity College Dublin, Dublin 2, Ireland}
\author{John Goold}
\email{gooldj@tcd.ie}
\affiliation{Department of Physics, Trinity College Dublin, Dublin 2, Ireland}
\author{Gabriel T. Landi}
\email{gtlandi@if.usp.br}
\affiliation{Instituto de F\'isica da Universidade de S\~ao Paulo,  05314-970 S\~ao Paulo, Brazil.}

\begin{abstract}

Thermodynamic uncertainty relations (TURs) place strict bounds on the fluctuations of  thermodynamic quantities in terms of the associated entropy production. In this work we identify the tightest (and saturable) matrix-valued TUR that can be derived from the exchange fluctuation theorems describing the statistics of heat and particle flow between multiple systems of arbitrary dimensions. Our result holds for both quantum and classical systems, undergoing general finite-time,  non-stationary processes. Moreover, it provides bounds not only for the variances, but also for the correlations between thermodynamic quantities. To demonstrate the relevance of TURs to the design of nanoscale machines, we consider the operation of a two-qubit SWAP engine undergoing an Otto cycle and show how our results can be used to place strict bounds on the correlations between heat and work.

\end{abstract}
\maketitle{}

%
%
%
%

{\bf \emph{Introduction} - }
Over the last  decades,  technological developments have  led to the creation of artificial meso- and nanoscopic heat engines~\cite{benenti2017fundamental, Hanggi2012RMP}, with applications ranging from nano-junction thermoelectrics~\cite{dubi2011colloquium} to quantum dots~\cite{josefsson2018quantum}. Understanding the fundamental principles ruling over the non-equilibrium physics of such devices is therefore one of nowadays most sought-after challenges.
One of the key features of these non-equilibrium processes is that they are always accompanied by an irreversible production of entropy. 
And as the systems become smaller, the fluctuations in the entropy production  become significant. 
This requires one to treat the entropy production $\Sigma$ as a random variable distributed according to a certain probability distribution  $P(\Sigma)$.
These distributions  satisfy a set of fundamental symmetry relations, known as Fluctuation Theorems (FT) \cite{Gallavotti1995b,Jarzynski1997a,Crooks1998,Piechocinska2000,Tasaki2000,Kurchan2000,Jarzynski2004a, Andrieux2009,Saito2008,Esposito2009,Campisi2011,Jarzynski2011a,Hanggi2015} which can generally be expressed as  $P(\Sigma)/\tilde{P}(-\Sigma) = e^{\Sigma}$, where $\tilde{P}(\Sigma)$ denotes the probability distribution of the time-reversed process.
FTs represent a refinement of the second law of thermodynamics, which at the stochastic level is recast in the form $\langle \Sigma \rangle \geq 0$. 
The additional information they carry, however, can also be used to characterize  systems arbitrarily far from equilibrium, which generated enormous interest  across many fields of research \cite{nitzan2003electron,galperin2007molecular, dubi2011colloquium, pekola2018thermodynamics, josefsson2018quantum, krinner2017two}.

More recently, another set of powerful results called \textit{Thermodynamic Uncertainty Relations} (TURs) have been discovered  \cite{Barato2015,Pietzonka2015,Pietzonka2017,Pietzonka2017a}. 
TURs impose strict restrictions on the fluctuations of thermodynamic currents  (e.g., heat, particles, etc.).
Letting $\mathcal{Q}$  denote any such \textit{integrated} current (net charge) exchanged during an out-of-equilibrium process over some generic time interval, the TURs bound the signal-to-noise ratio (SNR) of $\mathcal{Q}$ according to 
\begin{equation}\label{BS}
\frac{\var(\mathcal{Q})}{\langle \mathcal{Q} \rangle^2} \geq \frac{2}{\langle \Sigma \rangle},
\end{equation}
where $\var(\mathcal{Q}) = \langle \mathcal{Q}^2 \rangle - \langle \mathcal{Q} \rangle^2$ denotes the variance, $\langle \mathcal{Q}\rangle$  the average charge and $ \langle \Sigma \rangle$ the average entropy production.
Eq.~(\ref{BS}) expresses a tradeoff between process precision, quantified by the SNR, and dissipation, quantified through the entropy production. To reduce  fluctuations one must pay the inevitable price of dissipation.
This has important ramifications for the operation of microscopic autonomous engines \cite{Pietzonka2017} where fluctuations in the output power may be significant.  

TURs were originally discovered in the context of non-equilibrium steady-states of classical time-homogeoneous Markov jump-processes satisfying local detailed balance~\cite{Barato2015,Pietzonka2015,Pietzonka2017}. Further refinements and extensions have since then been found for finite-time processes~\cite{gingrich2016dissipation,Pietzonka2017a,Dechant2018}, periodically driven systems~\cite{BaratoNJP2018,Holubec2018PRL,proesmans2017discrete}, quantum systems in linear response~\cite{MacIeszczak2018} and using geometrical arguments based on the manifold of non-equilibrium steady-states~\cite{Guarnieri2019}. 

\begin{figure}
\centering
\includegraphics[width=0.45\textwidth]{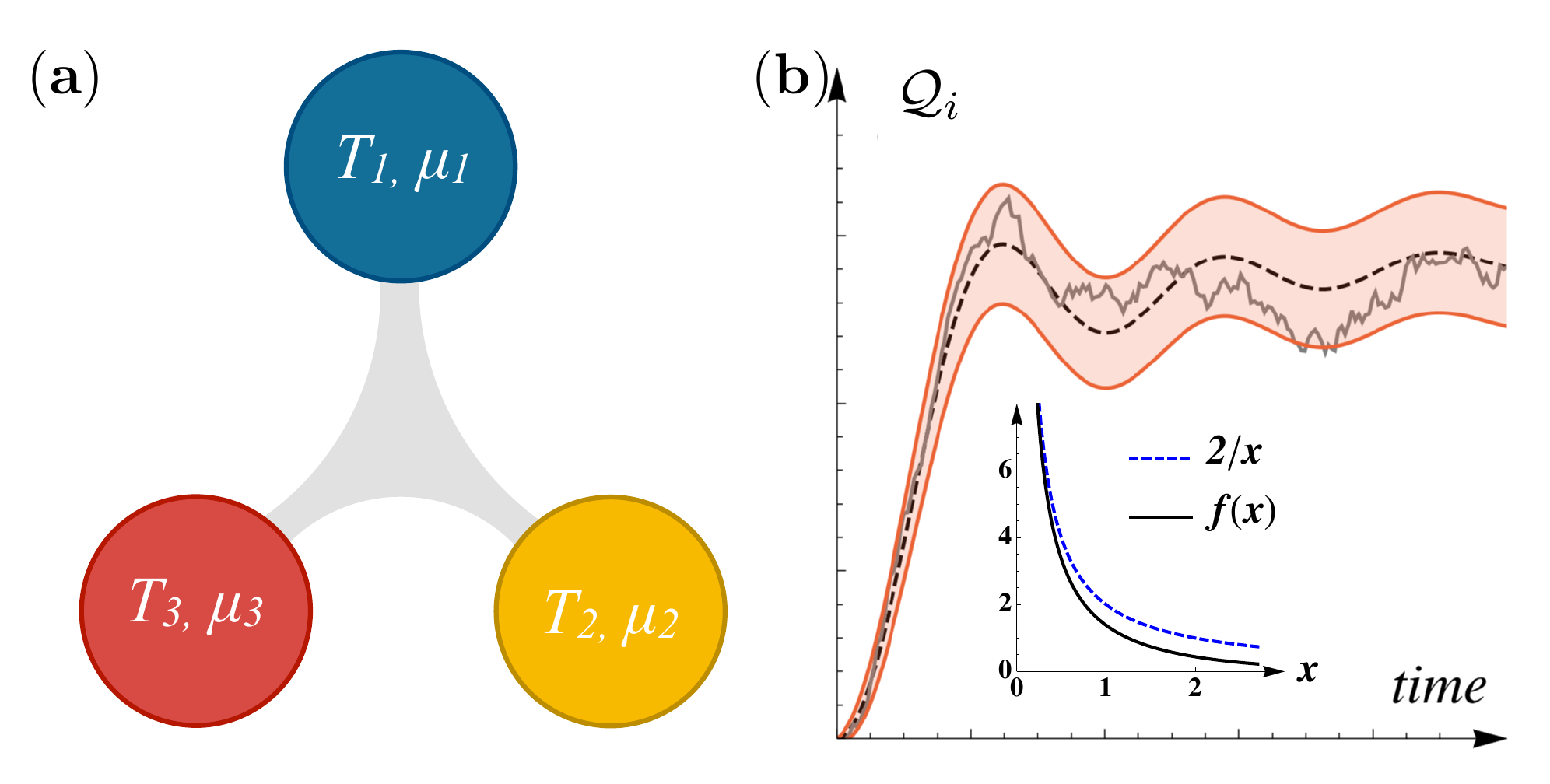}
\caption{\label{fig:drawing}
(a) Exchange Fluctuation Theorem scenario: A system consisting of $M$ (here $M = 3$) subsystems is allowed to interact by means of a unitary $U$.
As a result, the subsystems will exchange energy and particles, amounting to net transferred charges (integrated currents) of energy $\mathcal{Q}_{\mathcal{E}_i}$ and particles, $\mathcal{Q}_{\mathcal{N}_i}$. 
(b) For microscopic systems any generic $\mathcal{Q}_{i}$ will be a stochastic variable and fluctuate from one repetition of the experiment to the other, represented pictorially by the jagged gray curve.
This is to be contrasted with the average charge $\langle \mathcal{Q}_i \rangle$ shown as a dashed line. 
The fluctuations in $\mathcal{Q}_i$ are represented by the variance $\var(\mathcal{Q}_i)$, which we illustrate here by the red interval. 
Inset: a plot of the function $f(x)$ in the right-hand side of Eq.~(\ref{Timpa}), compared with the traditional bound $2/x$ that appears in Eq.~(\ref{BS}).
}
\end{figure}

A natural  question that emerges is whether TURs, being inequalities, can be viewed as a consequence of FTs, just like the second law $\langle \Sigma \rangle \geq 0$. 
Explorations in this direction began quite recently, starting with symmetric work protocols  \cite{Merhav2010,proesmans2017discrete,Hasegawa1} and subsequently generalized to include measurement feedback \cite{Hasegawa2,Potts2019}.
In this paper we derive a new type of saturable TUR for FTs stemming from heat and particle exchange between multiple systems (Fig.~\ref{fig:drawing}(a)).
This class of problems is particularly relevant, as it encompasses microscopic autonomous engines, which can be implemented in thermoelectric devices \cite{Hartmann2015,Thierschmann2015} and are now starting to be pursued in controlled quantum platforms \cite{Rosnagel2015,Klaers2017a,Peterson2018,Schmiegelow2018,Saunders2018,Gong2018,Maslennikov2017}.
A set of charges  $\mathcal{Q}_1, \ldots, \mathcal{Q}_n$  (energy, work, heat, particles, etc.) in this case satisfies the so-called  Exchange Fluctuation Theorems (EFTs) \cite{Jarzynski2004a,Andrieux2009,Saito2008}  (see also~\cite{Esposito2009,Campisi2011}) 
\begin{equation}\label{JW_FT}
\frac{P(\mathcal{Q}_1,\ldots,\mathcal{Q}_n)}{P(-\mathcal{Q}_1, \ldots, -\mathcal{Q}_n)} = e^{\sum\limits_i A_i \mathcal{Q}_i},
\end{equation}
where $A_i$ are thermodynamic affinities associated to each charge. 
The corresponding entropy production  is  $\Sigma = \sum_i A_i \mathcal{Q}_i$.

As our first main result, we show that the EFT~(\ref{JW_FT}) implies a generalized TUR for any charge $\mathcal{Q}_i$, of the form 
\begin{equation}\label{Timpa}
\frac{\var(\mathcal{Q}_i)}{\langle \mathcal{Q}_i \rangle^2} \geq f(\langle \Sigma \rangle),
\end{equation}
where $f(x) = \text{csch}^2(g(x/2))$, $\text{csch}(x)$ is the hyperbolic cosecant and $g(x)$ is the  function inverse of $x \tanh(x)$. 
We prove that this bound represents \textit{the tightest saturable} tradeoff bound for the SNR of any observable satisfying~\eqref{JW_FT},  given  $\langle \mathcal{Q}_i \rangle$. 
In fact, we are also able to provide an explicit form for the probability distribution $P(\mathcal{Q}_1,\ldots,\mathcal{Q}_n)$  saturating~(\ref{Timpa}).
This is to be compared, for instance, to the bounds derived in \cite{Merhav2010,proesmans2017discrete,Hasegawa1,Hasegawa2}, which are looser but cannot be saturated.
On the other hand, a series expansion of $f(x)$ around $x=0$ yields  $f(x) \simeq 2/x - 2/3$, so that for $\langle \Sigma \rangle \ll 1$ one recovers the bound~(\ref{BS}). 
The bound~(\ref{BS}), however, does not necessarily apply to all scenarios involving the exchange fluctuation theorem and can be violated. 
Our bound, on the other hand, is always looser than~(\ref{BS}) and always holds in any EFT scenario.

Our framework also allows us to go further and construct a matrix-valued TUR for the covariance matrix $\mathcal{C}_{ij} = \cov(\mathcal{Q}_i, \mathcal{Q}_j) = \langle \mathcal{Q}_i \mathcal{Q}_j \rangle - \langle \mathcal{Q}_i \rangle\langle \mathcal{Q}_j \rangle$ between different charges,  similar in spirit to Refs.~\cite{MacIeszczak2018,Dechant2018,Guarnieri2019}. 
In this case, the bound becomes
\begin{equation}\label{Timpa_matrix}
\mathcal{C} - f(\langle \Sigma \rangle) \bm{q} \bm{q}\trans \geq 0 ,
\end{equation} 
where $\bm{q} = (\langle \mathcal{Q}_1\rangle ,\ldots,\langle \mathcal{Q}_n \rangle)$ and the inequality is to be interpreted as a condition on the positive semi-definiteness of the matrix on the left-hand side. 
This bound therefore not only places restrictions on the fluctuations of currents, but also on their correlations. 

Eqs.~(\ref{Timpa}) and (\ref{Timpa_matrix}) are the main results of this paper. 
They hold for (i) quantum and classical systems of arbitrary dimensions  and (ii) undergoing arbitrarily  finite time processes far from equilibrium. The steady-state scenario is also contemplated as a particular case in which the systems become macroscopically large~\cite{Andrieux2009}. Below we start by reviewing the physical scenarios where our results are valid. We then  provide the details of the proof and discuss their physical consequences. To illustrate their usefulness, we then apply them to a two-qubit SWAP engine functioning as a nanoscale Otto cycle.

{\bf \emph{The Exchange Fluctuation Theorem (EFT) scenario} - }
We consider the scenario depicted in \ref{fig:drawing}(a) and studied in Refs.~\cite{Jarzynski2004a,Andrieux2009,Saito2008,Esposito2009,Campisi2011}. 
An arbitrary number $M$ of quantum systems are initially prepared in a factorized grand-canonical state $\rho = \prod_i Z_i^{-1} \exp\left[-\beta_i\left(\mathcal{H}_i - \mu_i \mathcal{N}_i\right)\right]$, where $\mathcal{H}_i, \mathcal{N}_i$ are the local Hamiltonians and particle number operators and $\beta_i, \mu_i$ the inverse temperature and chemical potential of the $i$-th subsystem \footnote{This also contemplates  the more typical scenario of a system coupled to multiple baths, but only provided the system starts in a thermal state. Otherwise, the currents will generally not obey a FT. For a more detailed discussion, c.f. Ref.~\cite{Campisi2011}.
}. The quantum systems are put in contact  at time $t=0$ up to a time $\tau$ by means of an arbitrary unitary $\hat{U}$ incorporating the effect of all interactions between the subsystems, as well as any possible external driving. The only assumption is that the external drives are time-symmetric, so that the unitary related to the time-reversed process is simply $\hat{U}^\dagger$. 
Classical systems can be treated in a similar way \cite{Jarzynski2004a}. 

As a result of this time-dependent protocol, the subsystems exchange both energy and particles with each other;
we  denote by $\mathcal{Q}_{\mathcal{E}_i} = \Delta \mathcal{E}_i$ and $\mathcal{Q}_{N_i} \equiv \Delta \mathcal{N}_i$ the integrated energy and particle currents during the time window $\left(0,\tau\right)$.
Following ~\cite{Andrieux2009,Campisi2011}, the full statistics of these quantities can be shown to satisfy the FT
\begin{equation}\label{energy_matter_FT}
\frac{P\left(\mathcal{Q}_{\mathcal{E}_1},\ldots ,\mathcal{Q}_{\mathcal{E}_M},\mathcal{Q}_{\mathcal{N}_1},\ldots ,\mathcal{Q}_{\mathcal{N}_M}\right)}{P\left(-\mathcal{Q}_{\mathcal{E}_1},\ldots ,-\mathcal{Q}_{\mathcal{E}_M},-\mathcal{Q}_{\mathcal{N}_1},\ldots ,-\mathcal{Q}_{\mathcal{N}_M}\right)} = e^{\sum_i\beta_i \left(\mathcal{Q}_{\mathcal{E}_i} - \mu_i\mathcal{Q}_{\mathcal{N}_i}\right)},
\end{equation}
which is of the form~(\ref{JW_FT}). 

Variations of Eq.~(\ref{energy_matter_FT}) may also be naturally constructed. 
Consider, for instance, the particularly relevant case of $M=2$ subsystems.
Particle conservation implies that it suffices to consider  the particle charge $\mathcal{Q}_{\mathcal{N}} = \Delta \mathcal{N}_2 = - \Delta \mathcal{N}_1$ and hence work only with $P(\mathcal{Q}_{\mathcal{E}_1}, \mathcal{Q}_{\mathcal{E}_2}, \mathcal{Q}_{\mathcal{N}})$. 
In addition, it may be of interest to change variables and use as thermodynamic quantities a heat charge  $\mathcal{Q}_{H} = -\Delta \mathcal{E}_{1}$ and a work charge $\mathcal{Q}_W = \Delta \mathcal{E}_1 + \Delta \mathcal{E}_2$.
The EFT for the joint distribution~\eqref{energy_matter_FT} then becomes~\cite{Sinitsyn2011,campisi2014JPhysA}
\begin{equation}\label{EFT2sys}
\frac{P(\mathcal{Q}_H,\mathcal{Q}_W,\mathcal{Q}_{\mathcal{N}})}{P(-\mathcal{Q}_H,-\mathcal{Q}_W,-\mathcal{Q}_{\mathcal{N}})} = e^{\delta \beta \; \mathcal{Q}_H + \beta_B \mathcal{Q}_W + \delta \beta\mu \; \mathcal{Q}_{\mathcal{N}}},
\end{equation}
where $\delta \beta = \beta_B - \beta_A$ and $\delta \beta \mu = \beta_A \mu_A - \beta_B \mu_B$ are the corresponding affinities. 
This result, as stated, does not assume any form of weak coupling or strict energy conservation (i.e., in general, $\Delta \mathcal{E}_1 \neq - \Delta \mathcal{E}_2$). 
But if that is the case, then no work is performed and it suffices to deal with $P(\mathcal{Q}_H, \mathcal{Q}_\mathcal{N})$. 

{\bf \emph{Derivation of the TUR} - }
We now turn to the derivation of our TUR bound. 
The starting point  is a general  joint probability distribution $P(\mathcal{Q}_1,\ldots,\mathcal{Q}_n)$ satisfying~(\ref{JW_FT}).
We first perform a change of variables to  $\Sigma = \sum_i A_i \mathcal{Q}_i$ and $Z = \sum_i z_i \mathcal{Q}_i$, where $z_i$ are a set of auxiliary variables.
The corresponding probability distribution
$P(\Sigma, Z) = \langle \delta(\Sigma - \sum_i A_i \mathcal{Q}_i) \delta (Z - \sum_i z_i \mathcal{Q}_i)\rangle$ will then have the same  symmetry as Eq.~(\ref{JW_FT}) \cite{Lahiri2015}. Namely, 
\begin{equation}\label{symmetry_Sigma_Z}
\frac{P(\Sigma, Z)}{P(-\Sigma, -Z)} = e^{\Sigma}.
\end{equation}
Our bound is now entirely based on the following  simple question:
\emph{for fixed $\langle \Sigma \rangle$ and $\langle Z \rangle$, what is the probability distribution $P(\Sigma, Z)$,  satisfying Eq.~(\ref{symmetry_Sigma_Z}), which has the smallest possible variance $\var(Z)$?} 
We call this the minimal distribution. Our main technical contribution can then be summarized by the following theorem:
\\

\noindent
{\bf Theorem} (\emph{``TUR de force''}){\bf .} \emph{For fixed finite $\langle \Sigma \rangle$ and $\langle Z \rangle$, 
the probability distribution $P(\Sigma,Z)$ satisfying~(\ref{symmetry_Sigma_Z}), with the smallest possible variance (the minimal distribution)  is the distribution 
\begin{multline}\label{minimal}
P_\text{min}(\Sigma,Z) =  \frac{1}{2\cosh(a/2)}\bigg\{e^{a/2} \delta\left(\Sigma-a\right)\delta\left(Z-b\right) \\
+ e^{-a/2}\delta\left(\Sigma+a\right)\delta\left(Z+b\right) \bigg\},
\end{multline}
where the values of $a$ and $b$ are fixed by $\langle \Sigma \rangle = a \tanh(a/2)$ and $\langle Z \rangle = b \tanh(a/2)$.}

The proof is given in the Supplemental Material~\cite{SupMat}.
We also note that a similar distribution also appears in Ref.~\cite{Merhav2010}.
For the minimal distribution~(\ref{minimal}), the variance of $Z$ is given by 
\begin{equation}\label{Z_saturation_minimal}
\var(Z)_\text{min} = \langle Z \rangle^2 f(\langle \Sigma \rangle),
\end{equation}
where $f(x)$ is the function discussed below Eq.~(\ref{Timpa}) and $\var(Z)_\text{min} $ is the variance of $Z$ calculated w.r.t. $P_\text{min}$ in Eq.~\eqref{minimal}. 
Proving that this distribution is minimal hence implies that 
\begin{equation}\label{bound_Z}
\var(Z) \geq f(\langle \Sigma\rangle) \langle Z \rangle^2,
\end{equation}
for any other probability distribution. 


{\bf \emph{Matrix-valued TUR} - }
We are now in the position to complete the derivation of our TUR. 
The  bound~(\ref{bound_Z}) holds for a general combination  $Z = \sum_i z_i \mathcal{Q}_i$ of the charges, with arbitrary parameters $z_i$. 
Let us then write $\langle Z \rangle = \sum_i z_i q_i$, where $q_i = \langle \mathcal{Q}_i \rangle$, and  $\var(Z) = \sum_{ij} \mathcal{C}_{ij} \;z_i z_j$, where $\mathcal{C}_{ij} = \cov(\mathcal{Q}_i, \mathcal{Q}_j)$. 
Eq.~(\ref{bound_Z}) can then also be written as 
\[
\bm{z}\trans \bigg( \mathcal{C} - f(\langle \Sigma \rangle) \bm{q} \bm{q}\trans\bigg) \bm{z} \geq 0.
\]
But since this must be true for any set of numbers $z_i$, it  follows that the matrix inside the parenthesis must itself be positive semidefinite. 
We therefore finally arrive at our main result in Eq.~(\ref{Timpa_matrix}); viz., 
$\mathcal{C} - f(\langle \Sigma \rangle) \bm{q} \bm{q}\trans \geq 0$. The positive semidefiniteness of this matrix implies that the diagonal entries must also be non-negative.
This then leads to Eq.~(\ref{Timpa}). 

In addition, a condition on the covariances may be obtained by using the fact that if $G$ is a positive semidefinite matrix, then $-\sqrt{G_{ii} G_{jj}} \leq G_{ij}  \leq \sqrt{G_{ii} G_{jj}}$. Applying this to Eq.~(\ref{Timpa_matrix})  immediately leads to 
\begin{equation}\label{bound_covariances}
f(\langle \Sigma \rangle) q_i q_j - M_{ij} \leq \mathcal{C}_{ij} \leq f(\langle \Sigma \rangle) q_i q_j + M_{ij},
\end{equation}
where $M^2_{ij} = (\var(\mathcal{Q}_i) - f(\langle \Sigma \rangle) q_i^2)(\var(\mathcal{Q}_j) - f (\langle \Sigma \rangle)q_j^2)$.  Beside their magnitude, a particularly relevant information is also contained in the \emph{sign} of the covariances $\mathcal{C}_{ij} = \cov(\mathcal{Q}_i, \mathcal{Q}_j)$.  
When $\mathcal{C}_{ij}$ is positive (negative),  values of $\mathcal{Q}_i$ above average imply values of $\mathcal{Q}_j$ above (below) average.

It is possible to find a simple criteria determining when  $\cov(\mathcal{Q}_i, \mathcal{Q}_j)$ will have a well defined sign (namely, the same as that of $q_i q_j$).  
This will occur whenever the lower and upper bounds in Eq.~\eqref{bound_covariances}  have the same sign, which amounts to checking whether $(f q_i q_j)^2 \geq M_{ij}^2$. 
Using the definition of $M_{ij}$ one then finds 
\begin{equation}\label{condition_signs_covariance}
\frac{q_i^2}{\var(\mathcal{Q}_i)} + \frac{q_j^2}{\var(\mathcal{Q}_j)} \geq \frac{1}{f(\langle \Sigma \rangle)}.
\end{equation}
If this inequality is satisfied, then it is guaranteed that $\mathrm{sign} \,\cov(\mathcal{Q}_i, \mathcal{Q}_j) = \mathrm{sign} \, q_i q_j$.

{\bf \emph{Application to a microscopic engine} - }
To illustrate our results, we consider the application of our bound to an  engine composed of  two qubits, with energy gaps $\epsilon_A$ and $\epsilon_B$, interacting by means of a SWAP unitary $\hat{U} = \frac{1}{2} (1 + \bf{\hat\sigma}_A \cdot \bf{\hat\sigma}_B)$, where $\hat\sigma_i$'s are the Pauli matrices~\cite{Campisi2015}. 
The non-resonant nature of the two qubits means that there will in general be a finite amount of work involved.
As shown in Ref.~\cite{Barra2015}, this work can physically be associated with the cost of turning the interaction between $A$ and $B$ on and off.
It is not necessary to specify precisely how this takes place, however. 
All we need is the form of the final unitary $\hat{U}$.

\begin{figure}
\centering
\includegraphics[width=0.44\textwidth]{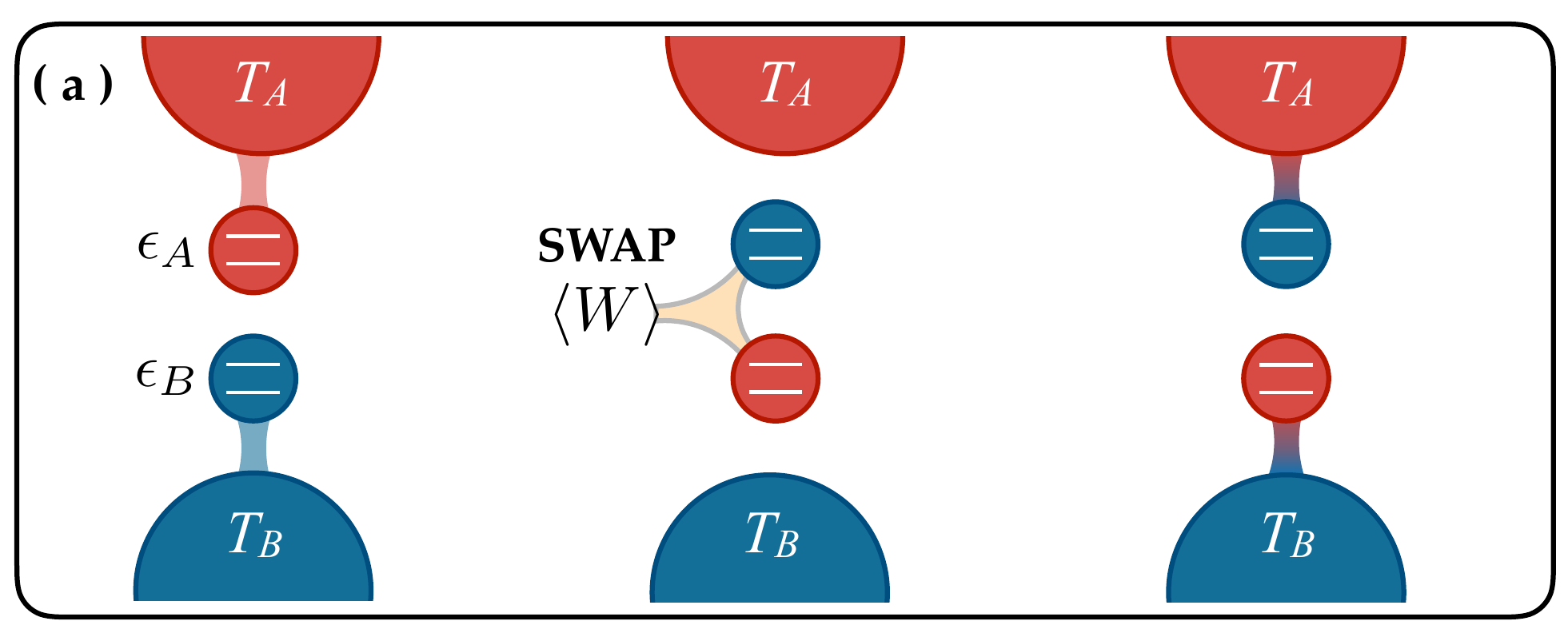}\\
\includegraphics[width=0.22\textwidth]{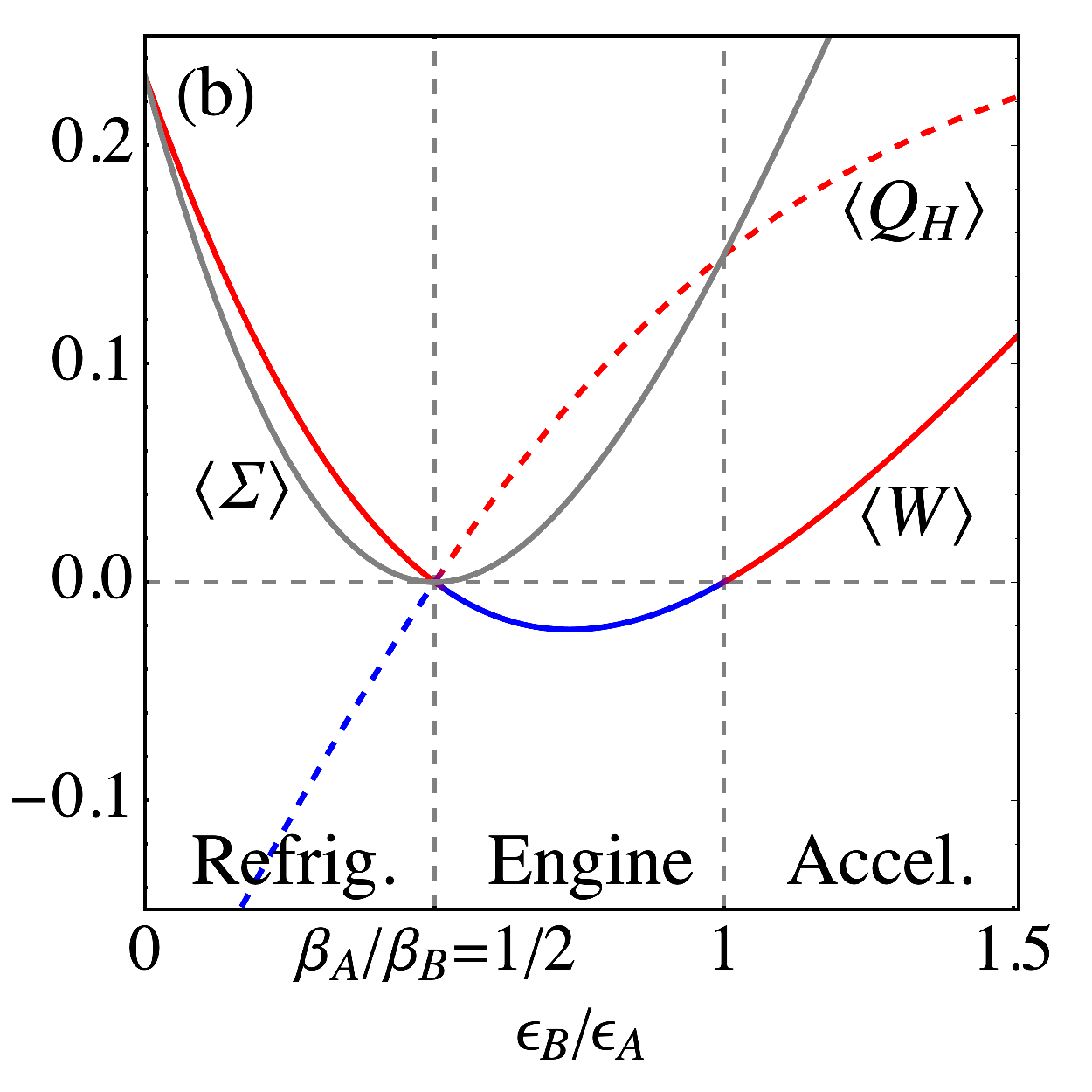}\quad
\includegraphics[width=0.22\textwidth]{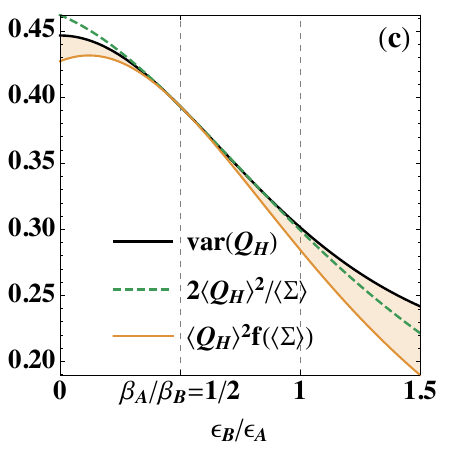}\\
\includegraphics[width=0.22\textwidth]{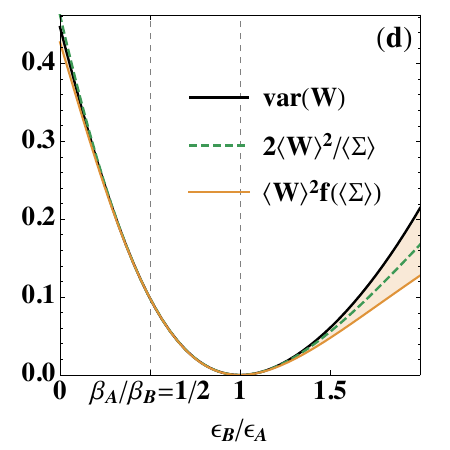}\quad
\includegraphics[width=0.22\textwidth]{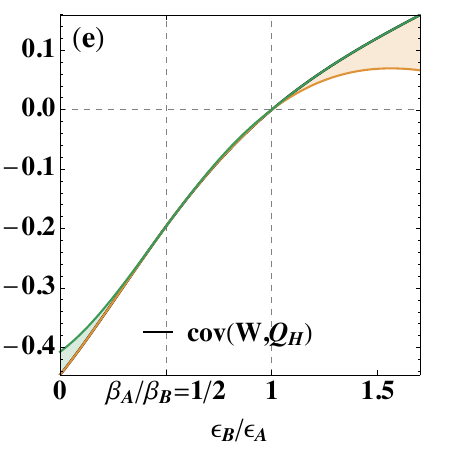}
\caption{\label{fig:engine}
Fluctuations of heat and work in a two-qubit SWAP engine.
(a) Schematic operation of the engine: 
Two qubits thermalize with two baths at different temperatures. 
Then they are uncoupled from the baths and allowed to interact with each other by means of a SWAP operation, which produces a certain amount of work $W$. 
Repeating this procedure sequentially allows the device to operate as either a refrigerator, an engine or an accelerator. 
(b) Averages of the work $\langle W \rangle$,  heat to the hot bath $\langle Q_H \rangle$ and entropy production $\langle \Sigma \rangle$ as a function of $\epsilon_B/\epsilon_A$ for $\beta_A/\beta_B = 1/2$. 
The different regimes of operation of the engine are separated by dashed vertical lines. 
(c) Fluctuations  in the heat to the hot bath $\var(Q_H)$. 
The orange line represents our bound~(\ref{Timpa}) and the green-dashed line represents the bound~(\ref{BS}), included for comparison. 
For small values of the  detuning $\epsilon_B/\epsilon_A$, (i.e. in the refrigerator regime), one can see that the bound~(\ref{BS}) is violated (the black line lies below the green dashed one) while~(\ref{Timpa}) is always valid. 
(d) Same but for the fluctuations in the work~$\var(W)$. 
(e) The correlations between heat and work, as measured by the covariance $\cov(W,Q)$. 
The two orange lines represent the bounds in Eq.~(\ref{bound_covariances}).
}
\end{figure}

After the qubits interact, one may reset their states by coupling them individually to two heat baths at different temperatures and allowing them to fully thermalize again (see Fig.~\ref{fig:engine}(a)).
Repeating the procedure sequentially then leads to a stroke-based engine operating at the Otto efficiency~\cite{DeChiara2018}. 
We assume $A$ is in contact with the hot bath, so $\beta_A < \beta_B$. 
The change in energy of qubit $A$ may thus be associated with the heat dumped into the hot reservoir, so we define $Q_H = - \Delta \mathcal{E}_A$. 
Similarly, the heat dumped to the cold reservoir is $Q_C = - \Delta \mathcal{E}_B$, whereas their mismatch is precisely the work, $W = - Q_H - Q_C = \Delta \mathcal{E}_A + \Delta \mathcal{E}_B$. 
The engine will thus be characterized by the stochastic variables $Q_H$ and $W$.
The corresponding probability distribution $P(Q_H,W)$, whose calculation details are presented in the Supplemental Material~\cite{SupMat}, will satisfy the EFT
\begin{equation}
\frac{P(Q_H,W)}{P(-Q_H,-W)} = e^{ (\beta_B - \beta_A) Q_H + \beta_B W},
\end{equation}
which is clearly of the form~(\ref{EFT2sys}), so that our basic framework applies.

Fig.~\ref{fig:engine}(b) shows $\langle W \rangle$, $\langle Q_H \rangle$  and $\langle \Sigma \rangle = (\beta_B-\beta_A) \langle Q_H \rangle + \beta_B \langle W \rangle$ as a function of $\epsilon_B/\epsilon_A$ with fixed $\beta_A/\beta_B = 1/2$. 
If $(\epsilon_B/\epsilon_A) < (\beta_A/\beta_B)$ the device operates as a refrigerator, consuming work from an external agent to make heat flow from the cold to the hot bath. 
Instead, if $(\beta_A/\beta_B) < (\epsilon_B/\epsilon_A) < 1$ it operates as a heat engine extracting useful work  ($\langle W \rangle < 0$). 
Finally, if $(\epsilon_B/\epsilon_A) > 1$ the device operates as an accelerator, consuming external work to increase the heat flow from hot to cold. 

In Fig.~\ref{fig:engine}(c) and (d) we present  results for the fluctuations of $Q_H$ and $W$ respectively. 
The results are compared with the bound~(\ref{Timpa}) as well as the bound~(\ref{BS}), included for comparison. 
As previously discussed, the bound~(\ref{BS}) can  be violated depending on the value of $\epsilon_B/\epsilon_A$. 
The bound~(\ref{Timpa}), on the other hand, is minimal and thus can never be violated.

Finally, in Fig.~\ref{fig:engine}(e) we present results for the covariance $\cov(W,Q_H)$. Studies on the correlations between thermodynamic quantities are still incipient \cite{MacIeszczak2018}. 
As can be seen in the image, in both the heat engine and the refrigerator regimes, the two quantities are negatively correlated, whereas for the accelerator they become positively correlated. 
The covariance in this case is bounded by the interval in Eq.~(\ref{bound_covariances}), which is represented by the two orange lines in Fig.~\ref{fig:engine}(e). 
For all parameters of this model, in the refrigerator regime the two bounds are always negative. 
Eq.~(\ref{condition_signs_covariance}), establishing the sign of $\cov(W,Q_H)$ is always satisfied only in the refrigerator regime. 
Thus, in this regime $\cov(W,Q_H)<0 $ and work and heat are always anti-correlated. 
In the other operation regimes,  such a general claim cannot be made.

{\bf \emph{Comparison with other TURs} - }
As discussed in the introduction, an expansion of Eq.~\eqref{Timpa} when $\langle \Sigma \rangle \ll 1$ leads to the original TUR Eq.~\eqref{BS}. These two bounds, however, must be compared with care, as they are derived for different physical scenarios.
The original TUR~(\ref{BS}) was  obtained for time-homogeneous Markovian jump-processes. 
Our bound, on the other hand, was derived assuming only the EFT. 
The two scenarios do not coincide. 
Indeed, as shown in  Fig.~\ref{fig:engine} (c), the bound~(\ref{BS}) can actually be violated in the EFT case. 
One situation for which the two scenarios could coincide is if the subsystems are macroscopically large. 
In this case there may exist  intermediate time intervals for which the exchange of energy will resemble that of a non-equilibrium steady-state \cite{Landi2014c}. 

It is also important to compare our bound with the one derived in Refs.~\cite{Merhav2010,proesmans2017discrete} which, translated into our notation, implies replacing the function $f(x)$ in Eq.~(\ref{Timpa}) with
\begin{equation}\label{HasegawaVu}
\frac{\var(\mathcal{Q}_i)}{\langle \mathcal{Q}_i \rangle^2} \geq \frac{2}{e^{\langle \Sigma\rangle} -1}.
\end{equation}
This bound is looser than both the original TUR~\eqref{BS} and our generalized TUR~(\ref{Timpa}). 
Moreover, relevant to the present letter, this bound was obtained by a different route than the one employed here, by means of a chain of inequalities~\cite{Hasegawa1}.
However, as we have just proved, the bound~(\ref{Timpa}) is the tightest possible bound and can only be saturated for a minimal distribution. 
As a consequence, the bound~(\ref{HasegawaVu}) can never be saturated. 
Indeed, in Ref.~\cite{Hasegawa2}, by the same authors, the bound~(\ref{HasegawaVu}) was replaced by a bound structurally identical to Eq.~(\ref{Timpa}).

Finally, we mention the connection with Ref.~\cite{Guarnieri2019}, where some of us have considered the non-equilibrium steady-state of a system connected to two infinite baths, a scenario where the original TUR~(\ref{BS}) can also be violated \cite{Agarwalla2018,Ptaszynski2018a}.
As this scenario does not satisfy an EFT,  in Ref.~\cite{Guarnieri2019} we approached the problem using the  Zubarev statistical ensemble, which allowed us to show that a TUR of the form~(\ref{BS}) also exists, but looser by a numerical factor. 
The two approaches therefore deal with different scenarios, but are both are motivated by the same drive to generalize TURs beyond their original formulation and into the quantum regime.

{\bf \emph{Conclusions} - }
In this Letter we have rigorously derived a new matrix-valued TUR solely as a consequence of EFT. This new tradeoff represents the tightest bound achievable on both the signal-to-noise ratio of any integrated current and for the covariance matrix between any pair of currents. Our derivation also allowed us to explicitly find the distribution saturating this ultimate bound. 
This result helps to answer in the affirmative the question of whether TURs, being inequalities, can also be viewed as a consequence of fluctuation theorems, much like the second law is obtained through Jensen's inequality. 
It hence places an important cornerstone in the direction of understanding and controlling non-equilibrium thermodynamic processes.

%


%

\emph{Acknowledgements - }
The authors acknowledge the International Institute of Physics in Natal, where part of this work took place, for both the financial support and the hospitality. 
The authors acknowledge fruitful discussions with D. Soares-Pinto, F. B. Brito, C. Boraschi, 
 P.-L. de Assis, F. Rouxinol, A. Amul, M. Terra Cunha and M. Motta.
GTL and JG acknowledge the hospitality of the Kauli Seadi school, where part of this work was developed.
GTL acknowledges the financial support from the University of S\~ao Paulo, the S\~ao Paulo Funding Agency FAPESP (grants 2017/50304-7, 2017/07973-5) and the Brazilian funding agency CNPq (Grant No. INCT-IQ 246569/2014-0). This work was supported by an SFI-Royal Society University Research Fellowship (J.G.) This project received funding from the European Research Council (ERC) under the European Union's Horizon 2020 research and innovation program (grant agreement No.~758403).

\bibliography{library}

\pagebreak
\widetext

\newpage 
\begin{center}
\vskip0.5cm
{\Large Supplemental Material}
\end{center}

\setcounter{equation}{0}
\setcounter{figure}{0}
\setcounter{table}{0}
\setcounter{page}{1}
\renewcommand{\theequation}{S\arabic{equation}}
\renewcommand{\thefigure}{S\arabic{figure}}
\renewcommand{\bibnumfmt}[1]{[S#1]}
\renewcommand{\citenumfont}[1]{S#1}

\newcommand{\prob}[1]{\mathrm{Pr}\left(#1\right)}
\newcommand{\EXP}{\mathds{E}}
\newcommand{\bra}[1]{\left< #1 \right|}
\newcommand{\ket}[1]{\left| #1 \right>}
\newcommand{\EX}[1]{\mathds{E}\left(#1\right)}
\newcommand{\condex}[2]{\mathds{E}\left(#1\,\middle|\, #2\right)}
\newcommand{\condvar}[2]{\mathrm{Var}\left(#1\,\middle|\, #2\right)}
\newcommand{\ep}[2]{E_{\substack{#1 \\ #2}}}
\newcommand{\detaux}[3]{\det\begin{bmatrix}#1 & #2 & #3\end{bmatrix}}
\newcommand{\detc}[3]{\det\left[\begin{array}{c|c|c}#1 & #2 & #3\end{array}\right]}
\newcommand{\av}[1]{\left<#1\right>}
\newcommand{\vvar}[1]{\var\left(#1\right)}

\theoremstyle{definition}
\theoremstyle{lemma}
\theoremstyle{theorem}

\newtheorem{lemma}{Lemma}[section]
\newtheorem{theorem}{Theorem}[section]
\newtheorem{definition}{Definition}[section]

This Supplementary Material is divided in two sections. 
In Sec.~I we provide details on the proof of Theorem 1 on the main text, which is the theorem required to demonstrate our generalized TUR. 
Then, in Sec.~II we provide some details on how to compute the thermodynamic quantities for the 2-qubit engine example summarized in Fig.~2 of the main text. 
An appendix is also included in the end, where we include more technical calculations postponed from Sec. I. 

\section{I. Proof of Theorem 1 of the main text. }

Here we prove the following result:  if we have a distribution $P(x,y)$ satisfying the symmetry relation
\begin{equation}
P(x,y) = P(-x,-y) e^{ x},
\label{eq:FT-2d}
\end{equation}
and with $\av{x}$ and $\av{y}$  finite, then
\begin{equation}
\vvar{y} \geq \av{y}^2 f(\av{x}),
\label{eq:bound}
\end{equation}
with $f(x) = \csch^2(\nicefrac{g(x)}{2})$, where $g^{-1}(x) = x\tanh{x}$.

Throughout the whole proof we call distributions obeying (\ref{eq:FT-2d}), FT distributions. 
Also note that eq (\ref{eq:FT-2d}) implies that $P$ can be written as
\begin{equation}
P(x,y) = S(x,y) e^{\nicefrac{x}{2}},
\label{eq:S}
\end{equation}
where $S$ is symmetric, which is a property we will use throughtout the proof to build the distributions.

The proof is organized in four parts, as follows:
\begin{enumerate}[{I}. A.]
\item  We first show that, if we restrict ourselves to FT distributions with at most 2 points in their support and find the minimum of $\vvar{y}$ given the values of $\av{x}$ and $\av{y}$, then this minimum is reached by exactly one distribution, which we will call the minimal distribution for these values. Furthermore, the equality in (\ref{eq:bound}) holds for the minimal distribution.
\item  We then show (by brute force) that if $P$ is a FT distribution with either 3 or 4 points in its support and $P'$ is the associated minimal distribution, then $\vvar{y}_{P'} \leq \vvar{y}_{P}$.
\item  With this result, we use the law of total variance to show that this implies the bound for the case of discrete FT distributions, with finite support.
\item  Finally, we show that given any FT distribution $\rho$, with finite $\av{x}, \av{y}$ and $\av{y^2}$, we can build a sequence of discrete FT distributions with finite support, whose averages $\av{x}, \av{y}$ and $\av{y^2}$ converge to the averages of $\rho$.
\end{enumerate}

\subsection{I.~A.~The minimal distribution}

We first recall that Eq.~(\ref{eq:FT-2d}) implies $\av{e^{-x}} = 1$, which by Jensen's inequality implies that $\av{x} \geq 0$, with the equality ($\av{x} = 0$) implying that $\vvar{x} = 0$. We want to find the minimal distribution, that is the distribution with at most 2 points in its support such that $\vvar{y}$ is minimum given $\av{x}$ and $\av{y}$. We also want to show that the minimal distributions obey the equality in (\ref{eq:bound}). Since the case $\av{x} = 0$ works differently, we will start with it.

\subsubsection{The $\av{x} = 0$ case}

If $\av{x} = 0$, then $x = 0$ everywhere in the support and Eq.~(\ref{eq:S}) implies that $P$ is symmetric. As a consequence, every FT distribution with $\av{x} = 0$ also has $\av{y}=0$. The minimum variance is trivially achieved by the distribution where $(0,0)$ is the only point in the support, as $\vvar{y} = 0$ in this case. Also, its trivial that this distribution obeys the equality in (\ref{eq:bound}).

\subsubsection{The $\av{x} > 0$ case}

Since the only possible FT distribution with a 1 point support has $\av{x} = 0$, then we only need to consider the case with 2 points. Eq (\ref{eq:S}) means that a FT distribution with support $\{(-a,-b), (a,b)\}$ must be of the form
\begin{equation}
P(x,y) = 
\left\{
\begin{array}{ll}
p e^{\nicefrac{a}{2}} & \mbox{for }(x,y) = (a,b),\\[0.2cm]
p e^{\nicefrac{-a}{2}} & \mbox{for }(x,y) = (-a,-b),
\end{array}
\right. 
\end{equation}
where $p = 1/(2 \cosh(a/2))$. 
The values of $a$ and $b$ are then fixed by the first moments according to:
\begin{equation}
\begin{array}{l}
\av{x} = a\tanh\left(\nicefrac{a}{2}\right), \\[0.2cm]
\av{y} = b\tanh\left(\nicefrac{a}{2}\right).
\end{array}
\end{equation}

Using the previously defined function $g$, defined such that $g(x)^{-1} = x \tanh(x)$, we have
\begin{equation}
\begin{array}{l}
a = 2 g(\nicefrac{\av{x}}{2}), \\[0.2cm]
b = \frac{a\av{y}}{\av{x}}.
\end{array}
\end{equation}
Since $y^2 = b^2$ for both points in the support, $\av{y^2} = b^2$, and we immediately arrive at 
\begin{IEEEeqnarray}{rCl}
\vvar{y} 
&=& \av{y}^2 f(\av{x})
\end{IEEEeqnarray}
So in this case there is exactly one distribution with the prescribed first moments and it obeys the equality in (\ref{eq:bound}).

\subsection{I.~B.~Supports with 3 or 4 points}

We now show that if a FT distribution has either 3 or 4 points in its support, then the associated minimal distribution (that is, the one with the same values of $\av{x}$ and $\av{y}$) has a smaller or equal $\vvar{y}$. By the same argument we gave in the last section, the case $\av{x} = 0$ is trivial, as it implies $\av{y} = 0$ and the corresponding minimal distribution has $\vvar{y} = 0$. As such we assume $\av{x}>0$. We also don't need to consider the case $\av{y} = 0$, because in this case the minimal distribution trivially has $\vvar{y} = 0$, so we will also assume $\av{y}\neq 0$

\subsubsection{The 3 point case}

In this case, the support is $\{(-a, -b), (0, 0), (a, b)\}$, with $a>0, b \neq 0$, and Eq.~(\ref{eq:S}) implies $P$ is of the form
\begin{equation}
P(x,y) = 
\left\{
\begin{array}{ll}
p e^{\nicefrac{a}{2}} & \mbox{for }(x,y) = (a,b),\\[0.2cm]
p e^{\nicefrac{-a}{2}} & \mbox{for }(x,y) = (-a,-b),\\[0.2cm]
q & \mbox{for }(x,y) = (0,0)
\end{array}
\right.
\end{equation}
Imposing normalization and the values of the first moments gives
\begin{equation}
\begin{array}{l}
q + 2p\cosh\left(\nicefrac{a}{2}\right) = 1,\\[0.2cm]
2ap\sinh\left(\nicefrac{a}{2}\right) = \av{x},\\[0.2cm]
2bp\sinh\left(\nicefrac{a}{2}\right) = \av{y},
\end{array}
\end{equation}
whose solution is
\begin{equation}
\begin{array}{l}
p = \frac{\av{x}}{2a\sinh\left(\nicefrac{a}{2}\right)}, \\[0.2cm]
q = 1 - \frac{\av{x}}{a \tanh\left(\nicefrac{a}{2}\right)}, \\[0.2cm]
b = \frac{a\av{y}}{\av{x}}.
\end{array}
\end{equation}
Furthermore, imposing $q \geq 0$, we have that $a$ is a free variable that obeys the bound
\begin{equation}
a \tanh\left(\nicefrac{a}{2}\right) \geq \av{x}.
\label{eq:a-upperbound}
\end{equation}
To prove that the minimal distribution has a smaller or equal $\vvar{y}$, we minimize
\begin{IEEEeqnarray}{rCl}
\nonumber
\av{y^2} 
&=& \frac{\av{y}^2}{\av{x}}\frac{a}{\tanh\left(\nicefrac{a}{2}\right)},
\end{IEEEeqnarray}
over $a$ subject to the bound (\ref{eq:a-upperbound}).
This function is monotonic increasing in $a$, meaning that the global minimum of $\av{y^2}$ happens when $a$ is the smallest possible. On the other hand, $a \tanh\left(\nicefrac{a}{2}\right)$ is monotonic, which implies that the smallest value of $a$ allowed by the bound (\ref{eq:a-upperbound}) is such that $a \tanh\left(\nicefrac{a}{2}\right) = \av{x}$ and hence $q = 0$.
The distribution thus has only 2 points in the support). 


\subsubsection{The 4 point case}

In this case, the support is $\{(-x_a, -y_a), (-x_b, -y_b), (x_a, y_a), (x_b, y_b)\}$ and the distribution is of the form
\begin{equation}\label{eq_4points_distribution}
P(x,y) = 
\left\{
\begin{array}{ll}
p_a e^{\nicefrac{x_a}{2}} & \mbox{for }(x,y) = (x_a, y_a),\\[0.2cm]
p_a e^{\nicefrac{-x_a}{2}} & \mbox{for }(x,y) =(-x_a, -y_a),\\[0.2cm]
p_b e^{\nicefrac{x_b}{2}} & \mbox{for }(x,y) = (x_b, y_b),\\[0.2cm]
p_b e^{\nicefrac{-x_b}{2}} & \mbox{for }(x,y) = (-x_b, -y_b).\\[0.2cm]
\end{array}
\right.
\end{equation}
The main idea is similar to the 3 point case. 
One  minimizes $\av{y^2}$ constrained by the normalization, the values of the first moments and the fact that $p_a, p_b \geq 0$, to find that the minima are such that either one of  $p_a$ or $p_b$ is 0 and hence there are actually only 2 points in the support. 
Since one now has more variables, however, the algebra becomes substantially more complicated.
In order not to hamper the readability of this Supplemental Material, we postpone this demonstration to Sec. III below.

%
%
\subsection{I.C.~Finite Supports}
%
%

Having proven the results for 3 and 4 points in the support, it is now straightforward to generalize the argument to distributions having finite support. 
We make the following definitions:
\begin{definition}[Symmetric Sets]
A set $S$ in $\mathbb{R}^2$ is said to be symmetric iff $(a,b) \in S \Leftrightarrow (-a, -b) \in S$.
\end{definition}

\begin{definition}[Reducible Sets]
A set $S$ is said to be reducible iff $S$ is finite, symmetric and such that for every FT distribution $P(x,y)$ that has $S$ as  support, there exists another FT distribution $P'(x,y)$ with support $S'$ such that
\[
\av{x}_{P'} = \av{x}_{P}\quad\quad \av{y}_{P'} = \av{y}_{P}\quad\quad \vvar{y}_{P'} \leq \vvar{y}_{P} \quad\quad\mbox{and}\quad |S'| < |S|.
\]
\end{definition}
\noindent
We now consider the following lemma: 
\begin{lemma}
If $S$ is reducible, then so is any finite symmetric set that contains it.
\end{lemma}
\noindent Before proving the lemma, we note that what we proved in the last section was that symmetric sets with 3 or 4 points are reducible ($P'$ in the definition we just gave would be the minimal distribution associated with $\av{x}_P$ and $\av{y}_P$). The reason why these are important is that any symmetric set with an even (odd) number of points has a symmetric subset with 4 (3) points in it, so proving the lemma implies that all finite symmetric sets with more than 2 points are reducible, from which it follows by induction that the minimal distribution associated with given values of $\av{x}$ and $\av{y}$ has the smallest possible value of $\vvar{y}$ out of all the FT distributions with finite support (as you can step by step remove points from the support, until you get to a support with at most 2 points, without ever increasing the variance), proving the bound for this class of distributions.

\emph{Proof of the Lemma:} This lemma follows from the laws of total expectation and variance. Suppose that $S$ is reducible and let $P(x,y)$ be a FT distribution with finite support $S\cup T$, where $S\cap T = \varnothing$. Note that this implies that $T$ is finite and symmetric. Using the law of total variance, we have
\begin{equation}
\vvar{y} = \EX{\condvar{y}{I_S}} + \vvar{\condex{y}{I_S}},
\end{equation}
where $I_S$ is the random variable that indicates $S$ (that is $I_S = 1$ if $(x,y)\in\,S$ and $I_S = 0$ otherwise). 

Let $A$ and $B$ be the distributions conditioned to $I_S = 1$ and $I_S = 0$ respectively and let $p>0$ be the probability of $S$ as an event. Since both $S$ and $T$ are symmetric, this means that $A$ (support $S$) and $B$ (support $T$) have the correct symmetry and hence they are both FT distributions. It follows that
\begin{equation}
\condex{y}{I_S} = 
\left\{
\begin{array}{ll}
\EX{y}_A & \mbox{ with probability }p, \\[0.2cm]
\EX{y}_B & \mbox{ with probability }1-p,
\end{array}
\right.
\label{eq:condex}
\end{equation}
and
\begin{equation}
\condvar{y}{I_S} = 
\left\{
\begin{array}{ll}
\vvar{y}_A & \mbox{ with probability }p, \\[0.2cm]
\vvar{y}_B & \mbox{ with probability }1-p.
\end{array}
\right.
\label{eq:condvar}
\end{equation}
By the law of total expectation, $\EX{\condex{y}{I_S}} = \EX{y}$, 
while Eqs.~(\ref{eq:condex}) and (\ref{eq:condvar}) imply that
\begin{equation}
\vvar{\condex{y}{I_S}} = p \EX{y}_A^2 + (1-p) \EX{y}_B^2 - \EX{y}^2.
\end{equation}
In addition,
\begin{equation}
\EX{\condvar{y}{I_S}} = p \vvar{y}_A + (1-p)\vvar{y}_B
\end{equation}
Combining these results then leads to,
\begin{equation}
\vvar{y}_P = p (\vvar{y}_A + \EX{y}_A^2) + (1-p)(\vvar{y}_B + \EX{y}_B^2) - \EX{y}^2_P.
\label{eq:var-total}
\end{equation}

Finally we use that $S$ is reducible. Since $A$ is a FT distribution with support $S$, there exists another FT distribution $C$, with support $R$, satisfying $|R|<|S|$ and such that
\[
\EX{x}_C = \EX{x}_A\quad\quad\EX{y}_C = \EX{y}_A\quad\mbox{and}\quad \vvar{y}_C \leq \vvar{y}_A.
\]
Consider then the distribution $P'$, describing the following random process:
\begin{itemize}
\item Draw a point $(x,y)$ from the distribution $P$.
\item If $(x,y) \notin S$, then $(x,y)$ is the result of the draw.
\item If $(x,y) \in S$, then draw $(x',y')$ from $C$. $(x', y')$ is the result of the draw.
\end{itemize}
This implies that $P' = pC+(1-p)B$, which is a FT distribution with support $R \cup T$. Furthermore $B$ is $P'$ conditioned to $I_S = 0$ and $C$ is $P'$ conditioned to $I_S = 1$. If $z = x$ or $y$, as drawn from $P'$, then
\begin{equation}
\condex{z}{I_S}_{P'} = 
\left\{
\begin{array}{ll}
\EX{z}_C & \mbox{ with probability }p, \\[0.2cm]
\EX{z}_B & \mbox{ with probability }1-p,
\end{array}
\right.
\label{eq:condex2}
\end{equation}
and from the law of total expectation one may immediately show that $\EX{z}_{P'}  = \EX{z}_P$, 
implying that $P$ and $P'$ have the same first moments. Doing the same reasoning we did to get to Eq.~(\ref{eq:var-total}) for $P'$ leads us to
\begin{IEEEeqnarray}{rCl}
\nonumber
\vvar{y}_{P'} &=& p (\vvar{y}_C + \EX{y}_C^2) + (1-p)(\vvar{y}_B + \EX{y}_B^2) - \EX{y}_{P'}^2	\\[0.2cm]
\nonumber
&=& p (\vvar{y}_C + \EX{y}_A^2) + (1-p)(\vvar{y}_B + \EX{y}_B^2) - \EX{y}_{P}^2	\\[0.2cm]
\nonumber
&=& \vvar{y}_P + p (\vvar{y}_C - \vvar{y}_A)   \\[0.2cm]
&\leq & \vvar{y}_P,
\end{IEEEeqnarray}
Thus, $P'$ has the same first moments as $P$ but a smaller or equal variance. The support of $P'$ is also smaller, since $|T\cup R| \leq |T| + |R| < |T| + |S| = |S\cup T|$. In other words, no matter what FT distribution $P$ is, we can build another FT distribution $P'$ with a smaller support and smaller or equal variance, so $S\cup T$ is reducible, finishing the proof.

%
%
\subsection{I.D.~Extension to general support}
%
%

We now finally bridge the gap between finite support and the general case using the following convergence theorem:

\begin{theorem}
For every FT distribution $\rho$, with support $\mathcal{S}\subseteq \mathds{R}^2$ and a finite family of differentiable functions $\{\phi_j\}$, such that $\av{|\phi_j|} < \infty\,\forall\,j$, there exists a sequence $\{D_n\}_{n=1}^{\infty}$ of discrete FT distributions with finite support, such that
\[
\lim_{n \rightarrow \infty} \av{\phi_j}_{D_n} = \av{\phi_j}_{\rho}.
\]
\end{theorem}
\noindent
This theorem completes the proof of our bound, because if there existed a distribution with infinite support, violating the bound
\[
\vvar{y} \geq \av{y}^2 f(\av{x}) \Leftrightarrow \av{y^2} \geq \av{y}^2 (f(\av{x}) + 1),
\]
then since $f$ is continuous and $x$, $y$, $y^2$ are differentiable then the sequence $\{D_n\}_{n=1}^{\infty}$ associated with the family $\{x, y, y^2\}$ that the theorem guarantees exists would eventually violate the bound too, leading to a contradiction because all $D_n$ would have finite support. An important detail is that this sequence is not necessarily weakly convergent to $\rho$ (which is not useful for us anyway, as weak convergence guarentees only convergence of the expectations of bounded continuous functions). Instead all this sequence guarantees is convergence of $\av{x}, \av{y}$ and $\av{y^2}$ (or any other finite family of differentiable functions).


\begin{proof}
We begin by noticing that if $P$ is a FT distribution
\begin{IEEEeqnarray}{rCl}
\nonumber
\int_{\mathds{R}^2} P(x,y) \phi(x,y) \ud x \ud y &=& \frac{1}{2}\int_{\mathds{R}^2} (P(x,y) \phi(x,y) + P(-x,-y) \phi(-x,-y)) \ud x \ud y	\\[0.2cm]
\nonumber
&=& \frac{1}{2}\int_{\mathds{R}^2} S(x,y)(\phi(x,y) e^{\nicefrac{x}{2}} + \phi(-x,-y)e^{\nicefrac{-x}{2}}) \ud x \ud y 	\\[0.2cm]
&=& \int_{\mathds{R}^2} T(x,y)\left(\frac{\phi(x,y) e^{\nicefrac{x}{2}} + \phi(-x,-y)e^{\nicefrac{-x}{2}}}{e^{\nicefrac{x}{2}} + e^{\nicefrac{-x}{2}}}\right) \ud x \ud y,
\label{eq:T-def}
\end{IEEEeqnarray}
where $T(x,y) = S(x,y)\cosh(\nicefrac{x}{2})$. Substituting $\phi = 1$ in (\ref{eq:T-def}) shows us that $T$ is a normalized distribution. Furthermore, $T$ is symmetric, there is a one to one correspondence between $T$ distributions and $P$ distributions and we can write any expected value of $P$ as an expected value of $T$:
\[
\av{\phi}_P = \av{\frac{\phi(x,y) e^{\nicefrac{x}{2}} + \phi(-x,-y)e^{\nicefrac{-x}{2}}}{e^{\nicefrac{x}{2}} + e^{\nicefrac{-x}{2}}}}_T.
\]
Since $\phi(x,y)$ being differentiable implies $\frac{\phi(x,y) e^{\nicefrac{x}{2}} + \phi(-x,-y)e^{\nicefrac{-x}{2}}}{e^{\nicefrac{x}{2}} + e^{\nicefrac{-x}{2}}}$ is also differentiable, then to prove the convergence of the expected values in the FT case we just need to prove the convergence of these deformed expected values in the symmetric case. In other words, proving the theorem for symmetric distributions (which will be much less cumbersome) automatically proves it for FT distributions.

Let $\mathcal{S}_{n}$ denote the square $[-n,n]^2$. We can break $\mathcal{S}_n$ into a symmetrical partition $\mathcal{S}_{n,m}$ with $(2m+1)^2$ disjoint squares. More precisely, $\mathcal{S}_{n,m} = \left\{I_{n,m}(k)\times I_{n,m}(q)\,\, \middle|\,\,k,q\in\{-n,1-n,\ldots, n\}\right\}$ with
\[
I_{n,m}(s) \equiv 
\left\{
\begin{array}{ll}
\left]\frac{2ns-n}{2m+1}, \frac{2ns+n}{2m+1}\right] & \mbox{if } s > 0, \\[0.2cm]
\left[\frac{2ns-n}{2m+1}, \frac{2ns+n}{2m+1}\right[ & \mbox{if } s < 0, \\[0.2cm]
\left[\frac{-n}{2m+1}, \frac{n}{2m+1}\right] & \mbox{if } s = 0.
\end{array}
\right.
\]
The elements of the partition can be identified by the integer vectors $\vec{k}$ that index them. For a given pair $(n,m)$ we can define $s^{n,m}_{\vec{k}}$ as the square defined by $\vec{k}$ in $\mathcal{S}_{n,m}$ and $c^{n,m}_{\vec{k}}$ as its center. Let $\rho(x,y)$ be a symmetric distribution, $\vec{x}$ be the random vector described by it and $\phi(x,y)$ be a differentiable function. We consider the non-normalized discrete distributions $\widetilde{D}_{n,m}$ defined over the centers of the squares in $\mathcal{S}_{n,m}$:
\[
\widetilde{D}_{n,m}\left(c^{n,m}_{\vec{k}}\right) = \prob{\vec{x}\in s^{n,m}_{\vec{k}}},
\]
and 0 elsewhere. Since $\mathcal{S}_{n,m}$ has $(2m+1)^2$ elements, then all $\widetilde{D}_{n,m}$ have finite support. Furthermore, since the partition $\mathcal{S}_{n,m}$ is symmetric, $\widetilde{D}_{n,m}$ is also symmetric. We now define
\[
\mathcal{E}_{n,m}[\phi] = \sum_{\vec{k}} \phi\left(c^{n,m}_{\vec{k}}\right)\widetilde{D}_{n,m}\left(c^{n,m}_{\vec{k}}\right),
\]
and
\[
\EXP_n[\phi] = \int_{\mathcal{S}_n} \phi(\vec{x})\rho(\vec{x})\ud \vec{x},
\]
which is to be understood as a Riemann-Lebesgue integral.
It follows that for every partition $\mathcal{S}_{n,m}$, we can use the mean value theorem to get
\begin{IEEEeqnarray}{rCl}
\nonumber
\EXP_n[\phi] &=& \sum_{\vec{k}}\int_{s^{n,m}_{\vec{k}}} \phi(\vec{x})\rho(\vec{x})\ud \vec{x}	\\[0.2cm]
\nonumber
&=& \sum_{\vec{k}} \phi\left(\vec{\mu}^{n,m}_{\vec{k}}\right)\int_{s^{n,m}_{\vec{k}}} \rho(\vec{x})\ud \vec{x} 	\\[0.2cm]
&=& \sum_{\vec{k}} \phi\left(\vec{\mu}^{n,m}_{\vec{k}}\right)\widetilde{D}_{n,m}\left(c^{n,m}_{\vec{k}}\right),
\end{IEEEeqnarray}
where $\vec{\mu}^{n,m}_{\vec{k}} \in s^{n,m}_{\vec{k}}$, so the difference $\Delta_{n,m}[\phi] = \mathcal{E}_{n,m}[\phi] - \EXP_n[\phi]$ is
\[
\Delta_{n,m}[\phi] = \sum_{\vec{k}} \widetilde{D}_{n,m}\left(c^{n,m}_{\vec{k}}\right) \left( \phi\left(c^{n,m}_{\vec{k}}\right) - \phi\left(\vec{\mu}^{n,m}_{\vec{k}}\right)\right).
\]
This implies that 
\[
\left| \Delta_{n,m}[\phi] \right| \leq \sum_{\vec{k}} \widetilde{D}_{n,m}\left(c^{n,m}_{\vec{k}}\right) \left| \phi\left(c^{n,m}_{\vec{k}}\right) - \phi\left(\vec{\mu}_{\vec{k}}^{n,m}\right)\right|.
\]
\noindent
Finally, $\phi$ is differentiable and $\mathcal{S}_n$ is compact, so $\phi$ is Lipschitz-continuous in $\mathcal{S}_n$, with associated constant $\kappa_n$ (that is $\left|f\phi(\vec{x}) - \phi(\vec{y})\right| \leq \kappa_n \left|\vec{x} - \vec{y}\right|\,\,\forall\,\,\vec{x}, \vec{y} \in \mathcal{S}_n$), implying
\begin{IEEEeqnarray}{rCl}
\nonumber
\left| \Delta_{n,m}[\phi] \right| &\leq & \kappa_n\sum_{\vec{k}} \widetilde{D}_{n,m}\left(C_{\vec{k}}\right) \frac{2n\sqrt{N}}{m}
=\frac{2n\kappa_n\mathcal{P}_n\sqrt{2}}{m}.
\label{eq:bound-lipschitz}
\end{IEEEeqnarray}
where $\mathcal{P}_n = \prob{\vec{x} \in \mathcal{S}_n}$. 
Here, the factor $2n\sqrt{N}/m$ represents the largest distance between points in $s^{n,m}_{\vec{k}}$

We now turn to the family $\{\phi_j\}$ of functions whose expectations we want to converge to the ones in $\rho$. For all of them $\left| \Delta_{n,m}[\phi_j] \right|$ is bounded according to (\ref{eq:bound-lipschitz}) (with each function having its Lipschitz constant $\kappa_{j,n}$). Furthermore, since the family is finite there exists a Lipschitz constant $\mathcal{K}_n = \max_j \{\kappa_{j,n}\}$ that works for all of them, implying
\[
\left| \Delta_{n,m}[\phi_j] \right| \leq \frac{2n\mathcal{K}_n\mathcal{P}_n\sqrt{2}}{m}\,\,\forall\,\, j .
\]
Whence,
\[
\left| \Delta_{n,m}[\phi_j] \right| \leq \epsilon \quad\mbox{if}\quad m > \frac{2n\mathcal{K}_n\mathcal{P}_n\sqrt{2}}{\epsilon}.
\]
Let $\{\epsilon_n\}$ be a sequence of positive numbers such that $\epsilon_n \rightarrow 0$ when $n\rightarrow \infty$ and let $m_n$ be an integer such that $m_n > \nicefrac{2n\mathcal{K}_n\mathcal{P}_n\sqrt{2}}{\epsilon_n}$. It follows that $\left| \Delta_{n,m_n}[\phi_j] \right| \leq \epsilon_n\,\,\forall\,\, j$. Let $n_0$ be the smallest $n$ such that $\mathcal{P}_n > 0$. Finally, we will prove that the sequence $\{D_n\}_{n=1}^{\infty}$ we are looking for can be built using $D_n = \nicefrac{\widetilde{D}_{n,m_n}}{\mathcal{P}_n}$ for $n > n_0$ (since we only care for the limits, $n \leq n_0$ is irrelevant). Since $\av{|\phi_j|} < \infty$, then the Lebesgue's Dominated Convergence Theorem implies that 
\begin{equation}
\av{\phi_j}_{\rho} = \int_{\mathds{R}^2} \phi_j(\vec{x})\rho(\vec{x})\ud \vec{x} = \lim_{n\rightarrow \infty} \int_{\mathcal{S}_n} \phi_j(\vec{x})\rho(\vec{x})\ud \vec{x},
\label{eq:limit}
\end{equation}
which is finite. On the other hand,

\begin{IEEEeqnarray}{rCl}
\nonumber
\left| \av{\phi_j}_{\rho} - \av{\phi_j}_{D_n} \right| & = & \left| \int_{\mathds{R}^2} \phi_j(\vec{x})\rho(\vec{x})\ud \vec{x} - \sum_{\vec{k}} \phi_j\left(c^{n,m_n}_{\vec{k}}\right)D_{n}\left(c^{n,m_n}_{\vec{k}}\right)\right| 	\\[0.2cm]
\nonumber
 & = & \left| \int_{\mathcal{S}_n^{\complement}} \phi_j(\vec{x})\rho(\vec{x})\ud \vec{x} + \int_{\mathcal{S}_n} \phi_j(\vec{x})\rho(\vec{x})\ud \vec{x} - \frac{1}{\mathcal{P}_n}\sum_{\vec{k}} \phi_j\left(c^{n,m_n}_{\vec{k}}\right)\widetilde{D}_{n,m_n}\left(c^{n,m_n}_{\vec{k}}\right)\right| 	\\[0.2cm]
\nonumber
& = & \left| \int_{\mathcal{S}_n^{\complement}} \phi_j(\vec{x})\rho(\vec{x})\ud \vec{x} + \left(1-\frac{1}{\mathcal{P}_n}\right)\int_{\mathcal{S}_n} \phi_j(\vec{x})\rho(\vec{x})\ud \vec{x} - \frac{\Delta_{n,m_n}[\phi_j]}{\mathcal{P}_n} \right|	\\[0.2cm]
\nonumber
& \leq & \left| \int_{\mathcal{S}_n^{\complement}} \phi_j(\vec{x})\rho(\vec{x})\ud \vec{x}\right| + \left(\frac{1-\mathcal{P}_n}{\mathcal{P}_n}\right)\int_{\mathcal{S}_n} \left|\phi_j(\vec{x})\right|\rho(\vec{x})\ud \vec{x} + \frac{\left|\Delta_{n,m_n}[\phi_j]\right|}{\mathcal{P}_n} \\[0.2cm]
& \leq & \left|\av{\phi_j}_{\rho} - \int_{\mathcal{S}_n} \phi_j(\vec{x})\rho(\vec{x})\ud \vec{x}\right| + \left(\frac{1-\mathcal{P}_n}{\mathcal{P}_n}\right)\av{\left|\phi_j(\vec{x})\right|} + \frac{\epsilon_n}{\mathcal{P}_n}.
\label{eq:end}
\end{IEEEeqnarray}
However, each of these three terms go to zero separately as $n \rightarrow \infty$: first, due to Eq.~(\ref{eq:limit}),
\[
\lim_{n \rightarrow \infty} \left|\av{\phi_j}_{\rho} - \int_{\mathcal{S}_n} \phi_j(\vec{x})\rho(\vec{x})\ud \vec{x}\right| = 0.
\]
Second, since  $\mathcal{P}_n \rightarrow 1$ as $n \rightarrow \infty$, 
\[
\lim_{n \rightarrow \infty} \left(\frac{1-\mathcal{P}_n}{\mathcal{P}_n}\right)\av{\left|\phi_j(\vec{x})\right|} = 0.
\]
And finally, by construction, 
\[
\lim_{n \rightarrow \infty} \frac{\epsilon_n}{\mathcal{P}_n} = 0
\]
This therefore implies that
\[
\lim_{n\rightarrow \infty} \left| \av{\phi_j}_{\rho} - \av{\phi_j}_{D_n} \right| = 0,  
\]
or, what is equivalent,
\begin{equation}
\lim_{n\rightarrow \infty} \av{\phi_j}_{D_n} = \av{\phi_j}_{\rho},
\end{equation}
which concludes our proof. 

\end{proof}

\section{II. Thermodynamics of the two-qubit SWAP engine}

In this section we provide additional details on the two-qubit engine discussed in Fig. 2 of the main text. 
We consider two qubits with Hamiltonians $H_i = - \epsilon_i \sigma_+^i \sigma_-^i$, where $i \in \{A,B\}$, prepared in thermal states $\rho_i = e^{-\beta_i H_i}/\tr(e^{-\beta_i H_i})$. 
The two qubits are then put to interact by means of SWAP interaction: 
\begin{equation}
U = \frac{1}{2} (1 + {\bf \sigma}^A \cdot {\bf \sigma}^B) = \begin{pmatrix}
1 & 0 & 0 & 0 \\ 
0 & 0 & 1 & 0 \\
0 & 1 & 0 & 0 \\
0 & 0 & 0 & 1 
\end{pmatrix}.
\end{equation}
The interesting aspect of using a SWAP unitary is that it does not require one to describe the detailed microscopic (and possibly time-dependent) interaction operator $V_{AB}(t)$. 
It suffices to understand the overall unitary generated by this interaction. 

Since the two qubits are in general not resonant, the SWAP usually performs a finite amount of work. 
For concreteness we assume $T_A > T_B$. 
The state of the two qubits are measured before and after the interaction, from which one can reconstruct $\Delta \mathcal{E}_A$ and $\Delta \mathcal{E}_B$. 
After this process, the two qubits are then allowed to thermalize once again with their respective baths. 
Hence, one can define the heat flowing to the hot bath as $Q_H = - \Delta \mathcal{E}_A$. 
Similarly, we define the heat flowing to the cold bath as $Q_C = - \Delta \mathcal{E}_B$. 
Their mismatch is the overall work that had to be performed, $W = \Delta \mathcal{E}_A + \Delta E_B = -Q_H - Q_C$. 

The cumulant generating function (CGF) of $Q_H$ and $W$ is then given by 
\begin{equation}\label{CGF}
\mathcal{C}(\lambda_H, \lambda_W) 
= \ln \langle e^{\lambda_H Q_H + \lambda_W W}\rangle
= \ln \tr\bigg\{ U^\dagger e^{-\lambda_H H_A} e^{\lambda_W (H_A + H_B)} U e^{\lambda_H H_A} e^{-\lambda_W (H_A + H_B)} (\rho_1\otimes \rho_2) \bigg\}.
\end{equation}
From the characteristic function one can compute all cumulants of $Q_H$ and $W$. 
In particular:
\begin{IEEEeqnarray}{rCl}
\langle Q_H \rangle &=& \frac{\partial \mathcal{C}}{\partial \lambda_H}, \\[0.2cm]
\langle W \rangle &=& \frac{\partial \mathcal{C}}{\partial \lambda_W}, \\[0.2cm]
\var(Q_H) &=& 2\frac{\partial^2 \mathcal{C}}{\partial \lambda_H^2}, \\[0.2cm]
\var(W) &=& 2\frac{\partial^2 \mathcal{C}}{\partial \lambda_W^2}, \\[0.2cm]
\cov(Q_H,W) &=& \frac{\partial^2 \mathcal{C}}{\partial \lambda_W\lambda_H}.
\end{IEEEeqnarray}
The entropy production in this case is
\begin{equation}
\sigma = (\beta_B - \beta_A) Q_H + \beta_B W.
\end{equation}

Carrying out the explicit computation of the CGF~(\ref{CGF}) for our problem leads to
\begin{equation}\label{supmat_CGF_example}
\mathcal{C}(\lambda_H, \lambda_W) = 
\ln \left\{\frac{e^{-\epsilon_A \left(\lambda_H+\lambda_W\right)-\epsilon_B \lambda_W} \left(e^{\epsilon_A \lambda _H+\epsilon_B \lambda_W}+e^{\epsilon_A \left(\beta_A+\lambda_W\right)}\right) \left(e^{\epsilon_A \lambda_H+\beta_B \epsilon_B+\epsilon_B \lambda_W}+e^{\epsilon_A \lambda_W}\right)}{\left(e^{\beta_A \epsilon_A}+1\right) \left(e^{\beta_B \epsilon_B}+1\right)}\right\}.
\end{equation}
The formulas for the first moments read
\begin{IEEEeqnarray}{rCl}
\langle Q_H \rangle &=& \epsilon_A \; \frac{e^{\beta_B \epsilon_B} - e^{\beta_A \epsilon_A}}{(e^{\beta_A \epsilon_A} + 1)(e^{\beta_B \epsilon_B} + 1)}, \\[0.2cm]
\langle W \rangle &=& (\epsilon_B-\epsilon_A) \; \frac{e^{\beta_B \epsilon_B} - e^{\beta_A \epsilon_A}}{(e^{\beta_A \epsilon_A} + 1)(e^{\beta_B \epsilon_B} + 1)}.
\end{IEEEeqnarray} 
From this one can directly verify that this system operators at the Otto efficiency:
\begin{equation}
\eta = - \frac{\langle W \rangle}{\langle Q_H \rangle} = 1- \frac{\epsilon_B}{\epsilon_A}.
\end{equation}
The expressions for the cumulants can be computed in a similar way, but are somewhat cumbersome and will thus not be displayed. 

The standard fluctuation theorem is contained in the identity
\begin{equation}\label{supmat_CGF_cond1}
\mathcal{C}(\lambda_H = \beta_A - \beta_B, \lambda_W = -\beta_B) = 0,
\end{equation}
which can be verified from Eq.~(\ref{supmat_CGF_example}). 
This, however, does not imply that the forward and backward distributions are the same, which is a stronger condition appearing in the exchange fluctuation theorems and which formed the basis for the present work (cf Eq.~(2)) of the main text.  

In addition to Eq.~(\ref{supmat_CGF_cond1}), distributions for which the forward and backward process are identical also satisfy the stronger condition, 
\begin{equation}\label{supmat_CGF_cond2}
\mathcal{C}(\beta_A - \beta_B - \lambda_H, -\beta_B -\lambda_W) = \mathcal{C}(\lambda_H, \lambda_W),
\end{equation}
which, again, one may verify for Eq.~(\ref{supmat_CGF_example}).
Eq.~(\ref{supmat_CGF_cond2}) immediately implies the detailed exchange fluctuation theorem
\begin{equation}
\frac{P(Q_H,W)}{P(-Q_H,-W)} = e^{ (\beta_B - \beta_A) Q_H + \beta_B W},
\end{equation}
which is Eq.~(12) of the main text. 

%
%
\section{Appendix:~Proof of minimality for a 4-point distribution}
%
%

Here we prove the minimality argument for the 4-point distribution~(\ref{eq_4points_distribution}). 
We once again impose our constraints
\begin{equation}
\begin{array}{l}
2p_a\cosh\left(\nicefrac{x_a}{2}\right) + 2p_b\cosh\left(\nicefrac{x_b}{2}\right) = 1,\\[0.2cm]
2 p_a x_a\sinh\left(\nicefrac{ x_a}{2}\right) + 2 p_b x_b\sinh\left(\nicefrac{ x_b}{2}\right) = \av{x},\\[0.2cm]
2 p_a y_a\sinh\left(\nicefrac{ x_a}{2}\right) + 2 p_b y_b\sinh\left(\nicefrac{ x_b}{2}\right) = \av{y}.
\end{array}
\label{eq:syst-4pts}
\end{equation}
The system (\ref{eq:syst-4pts}) can be slightly simplified if we define $q_a = 2p_a\cosh\left(\nicefrac{x_a}{2}\right)$ and $q_b = 2p_b\cosh\left(\nicefrac{x_b}{2}\right)$, as well as $\bar{x}$ and $\bar{y}$, such that $\bar{x}\tanh\left(\nicefrac{\bar{x}}{2}\right) = \av{x}$ and $\bar{y}\tanh\left(\nicefrac{\bar{x}}{2}\right) = \av{y}$:
\begin{equation}
\begin{array}{l}
q_a + q_b = 1,\\[0.2cm]
q_a x_a\tanh\left(\nicefrac{x_a}{2}\right) + q_b x_b\tanh\left(\nicefrac{x_b}{2}\right) = \bar{x}\tanh\left(\nicefrac{\bar{x}}{2}\right),\\[0.2cm]
q_a y_a\tanh\left(\nicefrac{x_a}{2}\right) + q_b y_b\tanh\left(\nicefrac{x_b}{2}\right)  = \bar{y}\tanh\left(\nicefrac{\bar{x}}{2}\right).
\end{array}
\label{eq:syst-4pts-simplified}
\end{equation}
Without loss of generality we have $0 \leq x_a \leq \bar{x} \leq x_b$ (because of the mean value theorem) and $\bar{x}, \bar{y} \neq 0$ (because we are assuming $\av{x}, \av{y} \neq 0$). In terms of these new variables the second moment becomes
\begin{IEEEeqnarray}{rCl}
\nonumber
\av{y^2} 
&=& \color{black}2p_a\color{black}y_a^2\color{black}\cosh\left(\nicefrac{x_a}{2}\right)\color{black} + \color{black}2p_b\color{black}y_b^2\color{black}\cosh\left(\nicefrac{x_b}{2}\right)		\\[0.2cm]
&=& \color{black}q_a\color{black}y_a^2 + \color{black}q_b\color{black}y_b^2.
\end{IEEEeqnarray}
Finaly, the minimal distribution associated with $\bar{x}$ and $\bar{y}$ is the one with support $\{(-\bar{x},-\bar{y}), (\bar{x},\bar{y})\}$, so we can reword what we want to prove as showing that any solution to (\ref{eq:syst-4pts-simplified}) with $q_a, q_b \geq 0$ is such that $\av{y^2} \geq \bar{y}^2$.

The first complication is that the nature of the solutions of the system (\ref{eq:syst-4pts-simplified}) change on certain cases, that we'll consider separately.


{\bf The case $x_a = x_b = \bar{x}\neq 0$:} The system (\ref{eq:syst-4pts-simplified}) can be simplified to
\begin{equation}
\begin{array}{l}
q_a + q_b = 1,\\[0.2cm]
q_a y_a + q_b y_b = \bar{y},\\[0.2cm]
q_a y_a^2 + q_b y_b^2 = \av{y^2},
\end{array}
\label{eq:syst-4pts-a-b}
\end{equation}
and $\av{y^2} \geq \bar{y}^2$ follows from Jensen's inequality, finishing the proof in this case.


{\bf The case  $y_a = c x_a$ and $y_b = c x_b$ ($c\neq 0$):} The third equation in the system (\ref{eq:syst-4pts-simplified}) is superfluous (note that this case can only happen if $\av{y} = c \av{x}$, which implies $\bar{y} = c \bar{x}$), leading to
\begin{equation}
q_a = \left(\frac{x_b\tanh\left(\nicefrac{x_b}{2}\right) - \bar{x}\tanh\left(\nicefrac{\bar{x}}{2}\right)}{x_b\tanh\left(\nicefrac{x_b}{2}\right) - x_a\tanh\left(\nicefrac{x_a}{2}\right)}\right),
\label{eq:q-4pts}
\end{equation}
\begin{equation}
q_b = \left(\frac{\bar{x}\tanh\left(\nicefrac{\bar{x}}{2}\right) - x_a\tanh\left(\nicefrac{x_a}{2}\right)}{x_b\tanh\left(\nicefrac{x_b}{2}\right) - x_a\tanh\left(\nicefrac{x_a}{2}\right)}\right),
\label{eq:p-4pts}
\end{equation}
and
\begin{IEEEeqnarray}{rCl}
\nonumber
\frac{\av{y^2}}{c^2} 
&=&	x_a^2 + \left(\bar{x}\tanh\left(\nicefrac{\bar{x}}{2}\right) - x_a\tanh\left(\nicefrac{x_a}{2}\right)\right)\left(\frac{x_b^2 - x_a^2}{x_b\tanh\left(\nicefrac{x_b}{2}\right) - x_a\tanh\left(\nicefrac{x_a}{2}\right)}\right).
\label{eq:x2}
\end{IEEEeqnarray}
The point is that this means $\av{y^2}$ is monotonic increasing with $x_b$, for $x_b > 0$. To see this, it suffices to show that the function
\begin{equation}
\frac{x_b\tanh\left(\nicefrac{x_b}{2}\right) - x_a\tanh\left(\nicefrac{x_a}{2}\right)}{x_b^2 - x_a^2},
\label{eq:secant-1}
\end{equation}
is decreasing in $x_b$. Since $x^2$ is increasing and invertible for $x > 0$, if we define $z_{a(b)} = x_{a(b)}^2$, then saying (\ref{eq:secant-1}) is decreasing with $x_b$, for $x_b \geq 0$, is the same as saying that
\begin{equation}
\frac{\sqrt{z_b}\tanh\left(\nicefrac{\sqrt{z_b}}{2}\right) - \sqrt{z_a}\tanh\left(\nicefrac{\sqrt{z_a}}{2}\right)}{z_b - z_a},
\end{equation}
is decreasing with $z_b$, for $z_a,z_b\geq 0$, which follows from $\sqrt{x}\tanh\left(\nicefrac{\sqrt{x}}{2}\right)$ being concave for $x > 0$. If $\av{y^2}$ is increasing with $x_b$, then its global minimum happens when $x_b$ is the smallest possible. From Eqs.~(\ref{eq:q-4pts}) and (\ref{eq:p-4pts}), we see that this means $x_b = \bar{x}$. But then $q_a = 0$ and the minimal distribution is the one that attains the minimum.


{\bf All other cases:} The general argument is at heart similar to the one used in the last case, but with a more complex algebra. Let $z = \bar{ x}y/\bar{ y}$. The system (\ref{eq:syst-4pts-simplified}) can be slightly rewritten as

\begin{equation}
\begin{array}{l}
q_a + q_b = 1,\\[0.2cm]
q_a x_a\tanh\left(\nicefrac{ x_a}{2}\right) + q_b x_b\tanh\left(\nicefrac{ x_b}{2}\right) = \bar{x}\tanh\left(\nicefrac{\bar{ x}}{2}\right),\\[0.2cm]
q_a z_a\tanh\left(\nicefrac{ x_a}{2}\right) + q_b z_b\tanh\left(\nicefrac{ x_b}{2}\right) = \bar{x}\tanh\left(\nicefrac{\bar{ x}}{2}\right),\\[0.2cm]
q_a z_a^2 + q_b z_b^2 = \av{z^2}.
\end{array}
\label{eq:syst-4pts-2}
\end{equation}
Solving the first 2 equations in the system (\ref{eq:syst-4pts-2}) we have again
\[
q_a = \frac{ x_b\tanh\left(\nicefrac{ x_b}{2}\right) - \bar{ x}\tanh\left(\nicefrac{\bar{ x}}{2}\right)}{ x_b\tanh\left(\nicefrac{ x_b}{2}\right) -  x_a\tanh\left(\nicefrac{ x_a}{2}\right)},
\]
\begin{equation}
q_b = \frac{\bar{ x}\tanh\left(\nicefrac{\bar{ x}}{2}\right) -  x_a\tanh\left(\nicefrac{ x_a}{2}\right)}{ x_b\tanh\left(\nicefrac{ x_b}{2}\right) -  x_a\tanh\left(\nicefrac{ x_a}{2}\right)},
\label{eq:p-2d}
\end{equation}
so $q_a, q_b > 0$ iff $ x_a < \bar{ x} <  x_b$.

Combining the second and third equations from the system (\ref{eq:syst-4pts-2}) leads to:
\begin{equation}
q_a(z_a -  x_a)\tanh\left(\nicefrac{ x_a}{2}\right) + q_b(z_b - x_b)\tanh\left(\nicefrac{ x_b}{2}\right) = 0,
\end{equation}
which thus implies that 
\begin{equation}
\frac{ z_b -  x_b}{ z_a -  x_a} = \frac{-q_a\tanh\left(\nicefrac{ x_a}{2}\right)}{q_b\tanh\left(\nicefrac{ x_b}{2}\right)} := \lambda, 	
\end{equation}
which can equivalently be written as 
\begin{equation}
 z_b =  x_b + \lambda( z_a -  x_a).
\label{eq:phi-2d}
\end{equation}

This result means one can specify $x_a, x_b$ and $z_a$. 
However,  the constraints of normalization and the prescribed moments  define $q_a$, $q_b$ and $ z_b$. 
What we want is then to find the values of $ x_a,  x_b$ and $ z_a$ which yield the smallest $ \av{ z^2}$. Substituting Eq.~(\ref{eq:phi-2d}) we have
\begin{IEEEeqnarray}{rCl}
\nonumber
\av{z^2} &=& q_a z_a^2 + q_b z_b^2	\\[0.2cm]
\nonumber
&=& z_a^2(q_a + q_b \lambda^2) + 2 z_a\lambda q_b( x_b - \lambda x_a) + q_b( x_b - \lambda x_a)^2	\\[0.2cm]
&=& A z_a^2 + 2B z_a + C.
\label{eq:moment-quad}
\end{IEEEeqnarray}
$A>0$, so we can readily minimize over $z_a$. The minimum happens at $z_a = \nicefrac{-B}{A}$, where we get $\av{ z^2} = E( x_a,  x_b)$, with
\begin{IEEEeqnarray}{rCl}
E( x_a,  x_b) &=& C - \frac{B^2}{A}	
= \frac{q_a q_b( x_b - \lambda x_a)^2}{q_a + \lambda^2q_b}.
\end{IEEEeqnarray}
Next we multiply by $q_b^2\tanh^2\left(\nicefrac{ x_b}{2}\right)$ in the numerator and denominator, then substitute $\lambda$, leading to
\begin{IEEEeqnarray}{rCl}
E( x_a,  x_b) 
&=& \frac{q_a q_b( x_b q_b \tanh\left(\nicefrac{ x_b}{2}\right) +  x_a q_a\tanh\left(\nicefrac{ x_a}{2}\right))^2}{q_aq_b^2\tanh^2\left(\nicefrac{ x_b}{2}\right) + q_bq_a^2\tanh^2\left(\nicefrac{ x_a}{2}\right)}.
\end{IEEEeqnarray}
We now note that the expected value $ \av{x}$ shows up in the numerator and cancel some terms out
\begin{IEEEeqnarray}{rCl}
\nonumber
E( x_a,  x_b) 
&=& \frac{\bar{ x}^2\tanh\left(\nicefrac{\bar{ x}}{2}\right)^2}{q_b\tanh^2\left(\nicefrac{ x_b}{2}\right) + q_a\tanh^2\left(\nicefrac{ x_a}{2}\right)}	\\[0.2cm]
&=& \frac{\bar{ x}^2\tanh\left(\nicefrac{\bar{ x}}{2}\right)^2}{\tanh^2\left(\nicefrac{ x_a}{2}\right) + q_b(\tanh^2\left(\nicefrac{ x_b}{2}\right) - \tanh^2\left(\nicefrac{ x_a}{2}\right))}.
\end{IEEEeqnarray}
Since $\bar{ x}$ is fixed, then minimizing $E$ over $ x_b$ (for a given $ x_a$) is equivalent to maximizing
\begin{equation}
\widetilde{E}( x_b;  x_a) = q_b(\tanh^2\left(\nicefrac{ x_b}{2}\right) - \tanh^2\left(\nicefrac{ x_a}{2}\right)),
\label{eq:E2-aux}
\end{equation}
over $ x_b$ for a given $ x_a$. We recall that $ x_a < \bar{ x} <  x_b$. What we want to show is that the $ x_b$ that maximizes Eq.~(\ref{eq:E2-aux}) is $\bar{ x}$ and for that it suffices to show that $\widetilde{E}( x_b;  x_a)$ is strictly decreasing in $ x_b$.

We substitute Eq.~(\ref{eq:p-2d}) to get $\widetilde{E}$ explicitly as a function of $ x_a$ and $ x_b$:
\begin{equation}
\frac{\widetilde{E}( x_a;  x_b)}{\bar{ x}\tanh\left(\nicefrac{\bar{ x}}{2}\right) -  x_a\tanh\left(\nicefrac{ x_a}{2}\right)} = \frac{\tanh^2\left(\nicefrac{ x_b}{2}\right) - \tanh^2\left(\nicefrac{ x_a}{2}\right)}{ x_b\tanh\left(\nicefrac{ x_b}{2}\right) -  x_a\tanh\left(\nicefrac{ x_a}{2}\right)}.
\end{equation}
Since $\bar{ x}$ and $ x_a$ are fixed and $\bar{ x} >  x_a$, then we just have to prove the right hand side is decreasing in $x_b$ for $x_b > x_a$. It suffices to prove that
\begin{equation}
\frac{ x_b\tanh(x_b) -  x_a\tanh(x_a)}{\tanh^2(x_b) - \tanh^2(x_a)},
\label{eq:inc1}
\end{equation}
is increasing. Since $\tanh^2(x)$ is monotonic increasing and invertible for $x>0$, if we define $w_{a(b)} = \tanh^2(x_{a(b)})$ then saying (\ref{eq:inc1}) is increasing is tantamount to saying that
\begin{equation}
\frac{ \arctanh(\sqrt{w_b})\tanh(\arctanh(\sqrt{w_b})) -  \arctanh(\sqrt{w_a})\tanh(\arctanh(\sqrt{w_a}))}{w_b-w_a} = \frac{\sqrt{w_b}\arctanh(\sqrt{w_b}) - \sqrt{w_a}\arctanh(\sqrt{w_a})}{w_b - w_a},
\label{eq:inc2}
\end{equation}
is increasing in $w_b$, for $0 < w_a, w_b < 1$, which follows from $\sqrt{x}\arctanh(\sqrt{x})$ being convex in this interval. This implies that for $\av{ z^2}$ (and hence $\av{ y^2}$) to be minimum, then $ x_b$ must be the smallest possible, implying $ x_b=\bar{ x}$, but then by Eq.~(\ref{eq:p-2d}) we have $q_a = 0$ meaning the minimum is attained by the minimal distribution.

\end{document}